\documentclass[english,aps,preprint]{revtex4}
\usepackage[T1]{fontenc}
\usepackage[latin1]{inputenc}
\usepackage{amsmath,graphicx}

\newcommand{\nn}{\nonumber}
\newcommand{\be}{\begin{equation}}
\newcommand{\ee}{\end{equation}}
\newcommand{\bea}{\begin{eqnarray}}
\newcommand{\eea}{\end{eqnarray}}

\makeatletter

\providecommand{\LyX}{L\kern-.1667em\lower.25em\hbox{Y}\kern-.125emX\@}

\usepackage{babel}
\makeatother
\begin{document}

\preprint{preprint version}

\title{Non Thermal Neutralino Production in Deflected Anomaly Mediation}

\author{A. M. Lionetto}

\email{lionetto@roma2.infn.it}

\affiliation{Department of Physics and INFN Roma Tor Vergata}

\begin{abstract}
We study the effects of a non thermal neutralino production, due to the
late decay in the early universe of a single modulus field, in the
context of the deflected anomaly mediated scenario. 
In the regime in which the average number of neutralino produced in each
modulus decaying process is $\bar{N}_{{\rm LSP}}\ll 1$ also models with 
a thermal relic density below WMAP data
became acceptable models. We find out that these models belong
to three different classes with the common feature that the low thermal
relic density is entirely due to coannihilation effects. 
The neutralino annihilation cross section for these classes of models is not
particularly high compared with the highest cross sections attainable
in the generic framework of the MSSM.
Hence the detection prospects either by direct or indirect WIMP search
experiments are not encouraging.     
\end{abstract}
\maketitle

\section{Introduction}


The latest WMAP data~\cite{Spergel:2006hy} 
determine in a very precise way the matter content of the Universe.
For the cold dark matter (CDM) component the WMAP result is
\be
\Omega_{CDM}h^2=0.110\pm 0.007
\label{eq:wmapdata}
\ee 
where $\Omega_{CDM}=\rho_{CDM}/\rho_c$ is the ratio between the CDM
density $\rho_{CDM}$ and the critical
density of the Universe $\rho_c=1.8791 h^2\cdot
10^{-29} {\rm g}\,{\rm cm}^{-3}$ with $h$ the Hubble constant in
units of $100\,{\rm km}\, {\rm s}^{-1}\, {\rm pc}^{-1}$.
Eq.~(\ref{eq:wmapdata}) is able to put a strong constraint on the
parameter space of a generic supersymmetric model. 
In fact the lightest supersymmetric particle (LSP), usually the
neutralino, is one of the best motivated cold dark matter
candidate (as long as the R-parity is conserved) and one can compute the
relic density of the lightest neutralino solving the following Boltzmann equation
\begin{equation}
\frac{dn_\chi}{dt}+3Hn_\chi=-\left<\sigma v\right>(n_\chi^{2}-n_{eq}^{2})
\label{eq:boltzmann}
\end{equation}
where $n_\chi$ is the numerical neutralino density, $H$ is the Hubble
constant, $\left<\sigma v\right>$ is the averaged thermal cross section times
the neutralino pair relative velocity while $n_{eq}$ is the Boltzmann equilibrium number density.
The physical picture described by this equation is simple and the
numerical solution is even straightforward until one takes into account
coannihilations with particles degenerate in mass with the lightest neutralino.
In a first approximation one can show~\cite{Jungman:1995df} that the following relation
holds
\be
\Omega^{{\rm thermal}} h^2\equiv \frac{m_\chi n_\chi}{\rho_c}\simeq \left<\sigma v\right>^{-1}
\label{eq:oh2thermal_sigmav}
\ee
where $m_\chi$ is the neutralino mass. With $\Omega^{{\rm
    thermal}} h^2$ we denote the solution of the standard Boltzmann equation~(\ref{eq:boltzmann}).
Typically when one compute eq.~(\ref{eq:boltzmann}) in the mSUGRA
framework, in which the LSP is a very pure bino in almost all the
parameter space,
one finds that the relic density easily exceeds the WMAP result~\cite{Baer:2002fv}.
The same situation happens in other supersymmetry breaking
scenarios.
Moreover the relation~(\ref{eq:oh2thermal_sigmav}) roughly implies
the existence of some upper bound on the annihilation cross section. 
Hence if we want to study models with cross section higher than that
allowed by the thermal relic density constraint we have to consider
scenarios in which the condition~(\ref{eq:oh2thermal_sigmav}) can be relaxed.

The relic density computation can be heavily affected 
in a non standard cosmological scenario which introduces a
 non thermal source of cold dark matter production. Among the different
scenarios the neutralino production from late decays of scalar
or moduli fields is an elegant and ``natural''
possibility~\cite{Moroi:1999zb,decaying-scalar}. 
In particular in supergravity and superstring theories, where in
general the
scalar potential has many flat directions, the presence of
moduli fields that acquire masses from supersymmetry breaking is a
common feature (for a review in the string theory context see~\cite{Silverstein:2004id}).
Moduli fields have important cosmological consequences although their couplings
are suppressed by inverse powers of the Planck mass. They dominate the
energy density in the early Universe and if the modulus
mass is sufficiently low (for example of the order of the electroweak scale) it can destroy the standard big-bang
nucleosynthesis scenario. This problem is usually termed as
cosmological moduli problem (see for example~\cite{Banks:1995dt} and
references therein).
One interesting possibility to avoid this problem was given
in~\cite{Moroi:1999zb}. In that scenario the modulus mass is setted by
the gravitino mass which is quite high in anomaly mediated models.
In this paper we will study the same solution applied in the context
of the deflected anomaly
mediation~\cite{Rattazzi:1999qg,Okada:2002mv,Cesarini:2006jp} in which
the tachyon problem of the minimal anomaly mediation is elegantly solved. In
particular we concentrate on~\cite{Cesarini:2006jp} where we performed
a detailed analysis of the neutralino sector of this scenario.

The outline of the paper is as follows: in sec.~\ref{sec:moduli} we
describe how the presence of a modulus field changes the standard
relic density computation in the framework of a generic supersymmetry
breaking scenario. In sec.~\ref{sec:nonthermalrelic} we apply the
results of the previous section to the deflected anomaly case. In
sec.~\ref{sec:detection} we study the detection prospects for this
scenario. The last section is devoted to the conclusions.
  
\section{Relic Density and Moduli Fields\label{sec:moduli}}
We closely follow the analysis of Moroi and
Randall~\cite{Moroi:1999zb} with a single modulus field $\phi$. 
We consider a system of coupled
Boltzmann equations describing the time evolution of the wimp number
density $n_\chi$, the modulus number density $n_\phi$ and the
radiation density $\rho_{rad}$:
\bea
\frac{dn_\chi}{dt}+3Hn_\chi & = & \bar{N}_{{\rm LSP}}\Gamma_\phi
n_\phi-\left<\sigma v\right>\left(n_\chi^{2}-n_{eq}^2\right)\nn\\
\frac{dn_\phi}{dt}+3Hn_\phi & = & -\Gamma_\phi n_\phi\nn\\
\frac{\rho_{rad}}{dt}+4H\rho_{rad} & = &(m_\phi-\bar{N}_{{\rm
    LSP}}m_\chi)\Gamma_\phi n_\phi+2m_\chi \left<\sigma v\right>\left(n_\chi^{2}-n_{eq}^2\right)
\label{eq:boltzmodulus}
\eea
where $\Gamma_\phi$ is the decay rate of the modulus field and
$\bar{N}_{{\rm LSP}}$ is the average number of wimp produced in each
modulus decaying process. 
These are the two crucial parameters in this description. 
The modulus field decay rate $\Gamma_\phi$ can be formally written as a sum
\be
\Gamma_\phi=\sum_i \Gamma_\phi^{(i)}
\ee
where each single decay rate $\Gamma_\phi^{(i)}$ is associated to one
of the possible interactions terms
of the lagrangian in which $\phi$ is coupled  to one of the fields in the neutralino sector, namely the two gauginos
 $\tilde{B}$, $\tilde{W}$
and the two higgsinos $\tilde{H}_1$ and $\tilde{H}_2$.
The index $i$ runs over all possible contributions. 
In the positive deflected anomaly scenario all the
terms described in appendix A of~\cite{Moroi:1999zb} are present and
so we decide to use the same useful parametrization
\be
\Gamma_\phi=\frac{1}{2\pi}\frac{m_\phi^3}{M_{\rm P}^2}
\ee
where $m_\phi^3$ is the modulus field mass and $M_{\rm P}\simeq 2.4\cdot
10^{18}$ GeV is the reduced 4-dimensional Planck mass. 
The naive expectation for $m_\phi$ is
\be
m_\phi\simeq m_{3/2}=F_\varphi
\label{eq:gravitino_modulus}
\ee 
where $m_{3/2}$ is the gravitino mass and $F_\varphi$ is the auxiliary
supergravity field vev that breaks supersymmetry in
the anomaly sector. The last equality holds both in the minimal
and in the deflected anomaly scenario. This relation
guarantees that the modulus mass is in general quite high.
The presence of the modulus field changes the thermal
history of the early universe~\cite{Kolb:1988aj}. 
In particular in order to preserve a successful description of the big-bang
nucleosynthesis we must require that the reheating temperature
associated to the modulus field decay
\be
T_{RH}\simeq 10\,{\rm MeV}\,\left(\frac{m_\phi}{100\,{\rm TeV}}\right)\cdot\left(\frac{\Lambda_{eff}}{M_{\rm P}}\right)^{-1}\cdot\left(\frac{g_*}{10.75}\right)
\label{eq:reheating}
\ee
be of order $T_{RH}\simeq 10$ MeV. In the previous relation $\Lambda_{eff}$ is the effective interaction scale of the
modulus and $g_*$ counts the total number of effectively massless
degrees of freedom.
If we impose this condition then $m_\phi\simeq 100$ TeV, 
assuming that $\Lambda_{eff}\simeq M_{\rm P}$ and $g_*=10.75$. 
Taking into account~(\ref{eq:gravitino_modulus}) this implies the
following condition on the auxiliary
supergravity field vev
\be
F_\varphi\ge 100\,{\rm TeV}
\ee
The other condition that has to be satisfied to have a source of non thermal
neutralino production is 
\be
T_{RH}<T_{dec}\simeq\frac{m_\chi}{30}
\label{eq:decouplingtemp}
\ee
where with $T_{dec}$ we have denoted the neutralino decoupling
temperature. By definition this is the temperature at which the
annihilation rate becomes smaller then the Hubble expansion rate.
The condition~(\ref{eq:decouplingtemp}) is necessary because in order to have
a neutralino production by the decay of $\phi$ this must happens far
away from 
the thermal and chemical equilibrium.
To compute the relic density one must numerically solve
 the system of equations~(\ref{eq:boltzmodulus}). 
However it is possible to give a
simple estimate~\cite{Moroi:1999zb} of the solution in two different regimes
\begin{equation}
\Omega_\chi h^2=\left\{
\begin{array}{ll}
\frac{3 m_\chi\Gamma_\phi}{2(2\pi^2/45)g_* T_{RH}^3 \left<\sigma
  v\right>}\frac{h^2}{\rho_c/s} & \qquad \bar{N}_{{\rm LSP}}\simeq 1\\
& \\
\frac{3 \bar{N}_{{\rm LSP}}\, m_\chi\Gamma_\phi^2 M_{\rm
    P}^2}{(2\pi^2/45)g_* T_{RH}^3 m_\phi}\frac{h^2}{\rho_c/s} &
 \qquad \bar{N}_{{\rm LSP}}\ll 1
\end{array}\right.
\label{eq:nonthermalrelic}
\end{equation}
where $\rho_c/s\simeq 3.6\cdot 10^{-9}\cdot h^2$ GeV is the critical
density to entropy ratio. 
A key observation about eq.~(\ref{eq:nonthermalrelic}) is that in the
case of very low number of LSP produced in each decay the relic
density is completely unrelated to the thermal averaged cross section $\left<\sigma
  v\right>$.

\section{Non Thermal Relic Density and Deflected Anomaly\label{sec:nonthermalrelic}}
We now apply the results of the previous section to the
deflected anomaly mediation case and in particular to the scenario
proposed in~\cite{Cesarini:2006jp}. This scenario differs from
minimal anomaly mediation for the presence of an
additional gauge mediated supersymmetry breaking sector 
that avoids tachyonic masses for the
sleptons (which do not transform under the $SU(3)$ gauge group). 
In this sector there is a gauge singlet chiral
superfield $X=\left(A_X,\Psi_X,F_X\right)$ which is directly coupled
to $N_f$ copies of messenger chiral superfields $\Phi_i$ and
${\bar{\Phi}}_i$, transforming under the fundamentals
and anti-fundamentals of the standard MSSM gauge groups (for a
comprehensive review see~\cite{Giudice:1998bp}).
The soft breaking sector of the model is specified, as we saw in the
previous section, in terms of the auxiliary
supergravity field vev $F_\varphi$ that breaks supersymmetry in
the anomaly sector inducing as well a breaking in the additional gauge
sector and in terms of the
ratio $f/m$, where $f\equiv \left<F_X\right>$ is
the auxiliary gauge singlet field vev and $m$ is the messenger mass scale.   

 In~\cite{Cesarini:2006jp} we
computed the standard thermal relic density without any
non thermal contribution using DarkSUSY~\cite{Gondolo:2004sc} taking
into account all the coannihilations. In that case the
auxiliary supergravity field
can be as low as $F_\varphi\simeq 10^2$ GeV. 
The WMAP allowed models are those in
which the neutralino is a very pure
bino or a
very pure higgsino depending on the number of messenger fields
$N_f$ (for $N_f\ge 2$ a higgsino branch open up and this branch
becomes greater for $N_f\gtrsim 5$). 
We declare that a neutralino is gaugino-like (in particular in our
case bino-like) if $Z_g>0.9$ while is higgsino-like when $Z_g<0.1$,
where $Z_g$ is the usually defined gaugino fraction. 
In all the intermediate cases we denote the neutralino as mixed-like.

In the new scenario with a non thermal contribution in order to satisfy the constraint on the reheating
temperature (see eq.~(\ref{eq:reheating})) we must impose 
\be
F_\varphi\ge 10^5\, {\rm GeV}
\label{eq:fphiconstraint}
\ee
that cuts out all the lower part of the parameter space explored
in~\cite{Cesarini:2006jp}.
The condition~(\ref{eq:decouplingtemp}) on the decoupling temperature
turns out to be always satisfied.
Moreover we want to add another constraint to select models with the
highest possible annihilation cross sections.
This condition, having in mind the relation~(\ref{eq:oh2thermal_sigmav}), can be translated as an upper bound on the thermal relic
density
\be
\Omega_\chi^{{\rm thermal}} h^2<0.09
\label{eq.thermalbound}
\ee
where $0.09$ is the $3\sigma$ lower bound of the WMAP
result~(\ref{eq:wmapdata}). This allows us to use the results of the
detailed scan performed in~\cite{Cesarini:2006jp}.

\begin{figure}[t]
  \begin{center}
    \includegraphics[scale=0.5]{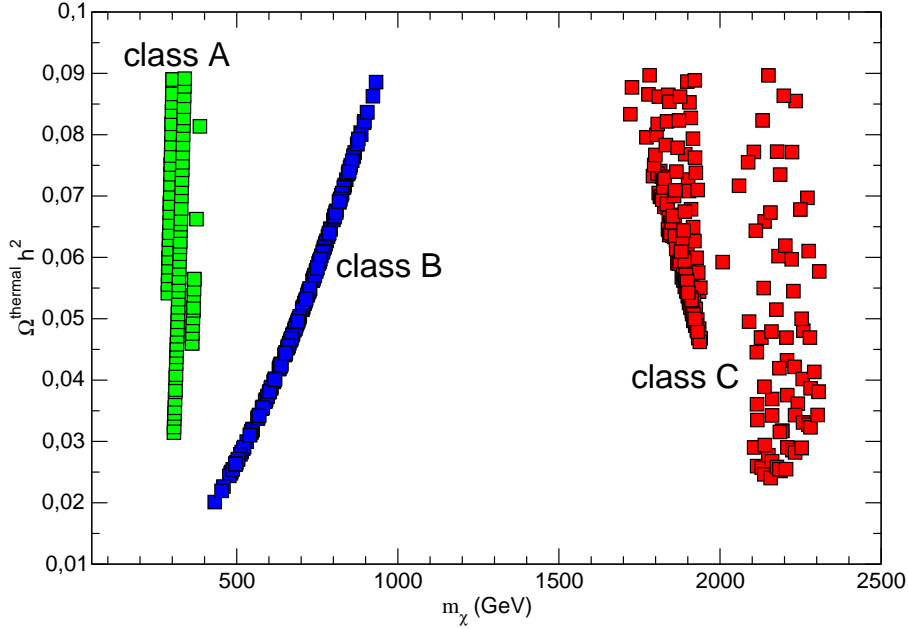}
    \caption{Allowed models with $\Omega_\chi^{{\rm
    thermal}}h^2 <0.09$ as a function of the neutralino mass $m_\chi$}
    \label{fig:oh2therm_vs_mchi}
  \end{center}
\end{figure}

Applying both constraints~(\ref{eq:fphiconstraint})
and~(\ref{eq.thermalbound}) it is possible to rule out the scenario with a high messenger
mass scale $m\gtrsim 10^{8}$ GeV. In fact in this case there is no higgsino branch and the neutralino is a very pure bino 
with a very low annihilation cross section.
For a lower messenger
mass scale $m\simeq 10^5$ GeV there are essentially three class of
models surviving denoted by A, B and C. For $m< 10^5$ GeV almost all the parameter
space is excluded either due to the presence of tachyons or because
there is no electroweak symmetry breaking or because the
neutralino is not the LSP. The result for $m\simeq
10^5$ GeV is conveniently summarized in
fig.~\ref{fig:oh2therm_vs_mchi} where we show in the plane
($\Omega_\chi^{{\rm thermal}} h^2$, $m_\chi$) the three class of
allowed models.
The plot was achieved applying the two bounds~(\ref{eq:fphiconstraint})
and~(\ref{eq.thermalbound}) on the entire parameter space. 
In the first class of models (class A) the
neutralino is a very pure bino with $200\;{\rm GeV}\lesssim m_\chi\lesssim 400$
GeV. These models are present for $N_f\le 5$. In the second class of
models (class B) the neutralino is a very pure higgsino with $400\;{\rm GeV}\lesssim
m_\chi\lesssim 900$ GeV. These models are always present as long as the
higgsino branch is open, i.e. $N_f>2$. 
Class B is the one that resembles more the scenario considered
in~\cite{Moroi:1999zb} where the neutralino is essentially a wino. 
The wino annihilates in a very efficient way
 into $W^+ W^-$ and $Z^0 Z^0$ (when this channels are kinematically opened)
so the thermal relic density is usually very low. 
In an analogous way the higgsino
has a quite efficient annihilation into $W^+ W^-$ and $Z^0 Z^0$ although
the higgsino cross section is lower than the wino one of about a factor
$10^{-2}$.
For class B models the non
thermal modulus contribution could be extremely important to enhance
the relic density. 
In the last class of models (class C) the
neutralino has a gaugino fraction $Z_g>0.1$ so it can be either mixed or bino like. In
both cases it has a very high mass of order $m_\chi>1.5$ TeV.
In table~\ref{tb:spectrum} we show the mass spectrum for three sample models
(one of each kind A, B or C). The first five rows define the model.
The next three rows contain the masses of the lightest
neutralino $m_{\tilde{\chi}}$ of the next to lightest neutralino
$m_{\tilde{\chi}_2}$ and of the lightest chargino
$m_{\tilde{W}_1}$. The next six rows show the mass spectrum of the
squark and slepton sector: the gluino mass $m_{\tilde{g}}$, the stop
mass $m_{\tilde{t}}$, an average mass of all the other squarks
$m_{\tilde{q}}$ (they are near degenerate), the masses of the left and
right selectrons, $m_{\tilde{e}_L}$ and $m_{\tilde{e}_R}$ respectively,
and finally the mass of the lightest stau $m_{\tilde{\tau}_1}$.
The last two rows show the mass spectrum of the Higgs sector: the mass
of the lightest Higgs $m_{h^0}$ and that of the CP-odd Higgs $m_{A^0}$
(quite degenerate with the remaining mass eigenstates).
\begin{table}[t]
\begin{tabular}{l|c|c|c}
& class A & class B & class C \\
\hline
$F_\varphi$ (GeV) & $10^5$ & $1.2\cdot 10^5$ &  $2\cdot 10^5$\\
$F_X/m$ (GeV) & $1.9\cdot 10^5$ & $4.9\cdot 10^5$ & $9.3\cdot 10^5$\\
$N_f$ & $5$ & $10$ & $4$ \\
$\tan\beta$ & $50$ & $10$ & $50$ \\
$sign(\mu)$ & $+1$ & $+1$ & $+1$ \\
\hline
$m_{\tilde{\chi}}$ & $290$ GeV & $705$ GeV & $2.1$ TeV\\
$m_{\tilde{\chi}_2}$ & $912$ GeV & $707$ GeV & $2.2$ TeV\\
$m_{\tilde{W}_1}$ & $912$ GeV & $706$ GeV & $2.2$ TeV\\
\hline
$m_{\tilde{g}}$ & $4.9$ TeV & $23$ TeV & $20$ TeV \\
$m_{\tilde{q}}$ & $4.2$ TeV & $15$ TeV & $17$ TeV\\
$m_{\tilde{t}}$ & $4$ TeV & $14$ TeV & $16$ TeV\\
$m_{\tilde{e}_L}$ & $1.2$ TeV & $5.4$ TeV & $6$ TeV\\
$m_{\tilde{e}_R}$ & $430$ GeV & $2.5$ TeV & $2.9$ TeV\\
$m_{\tilde{\tau}_1}$ & $294$ GeV & $2.5$ TeV & $2.5$ TeV\\ 
\hline
$m_{h^0}$ & $122$ GeV & $123$ GeV & $124$ GeV\\
$m_{A^0}$ & $1.4$ TeV & $5.2$ TeV & $4.4$ TeV\\
\end{tabular}
\caption{Mass spectrum for three benchmark models. See the text for
  the notation.}
\label{tb:spectrum}
\end{table}

\begin{figure}[t]
  \begin{center}
    \includegraphics{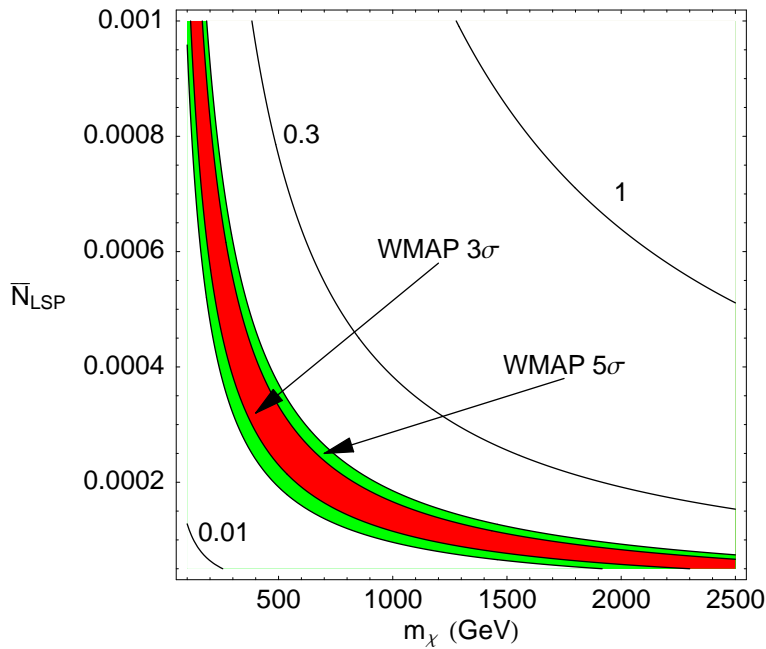}
    \caption{Contour plot of $\Omega_\chi h^2$ in the plane ($m_\chi$,
      $\bar{N}_{{\rm LSP}}$). The red (dark shaded) region corresponds
      to the $3\sigma$ WMAP result $0.09\le\Omega_\chi h^2\le 0.13$
      while the green (light shaded) region corresponds
      to the $5\sigma$ WMAP result $0.07\le\Omega_\chi h^2\le 0.14$.}
    \label{fig:nlsp_vs_mchi}
  \end{center}
\end{figure}

As can be seen the scalar sector tends to be heavy with the exception
of sleptons for class A models. In this case it is worth noting the
degeneracy between the lightest neutralino and the the lightest stau $\tilde{\tau}$.
For models B and C the key feature is the near
degeneracy between the lightest neutralino and the lightest chargino
$\tilde{W}_1$ and the next to lightest neutralino $\tilde{\chi}_2$.
The thermal relic density of all these models is shown in fig.~\ref{fig:oh2therm_vs_mchi}. As can be seen there is a lower bound $\Omega_\chi^{{\rm
    thermal}}h^2>0.02$.
 Hence there are no models with an
extremely low annihilation cross section and this implies (see eq.~(\ref{eq:nonthermalrelic})) that a relic density in the WMAP
range could only be obtained in the regime $\bar{N}_{{\rm LSP}}\ll 1$.
In fact, as we will see in more detail in the next section,
 for all models a thermal relic density below WMAP data is entirely due to coannihilation effects.   
In fig.~\ref{fig:nlsp_vs_mchi} we show the allowed values of $\bar{N}_{{\rm LSP}}$
for each value of the LSP mass $m_\chi$.
As quoted in~\cite{Moroi:1999zb} the exact value of $\bar{N}_{{\rm
    LSP}}$ is extremely model dependent. It relies on the
precise structure of the operators through which the modulus field can
decay.
In general the range of the allowed values is
\be
10^{-3}\lesssim\bar{N}_{{\rm LSP}}\lesssim 10^{-4}
\ee
Such range guarantees that all models fulfill the WMAP constraints at
least at a $5\sigma$ level.  

\section{Detection Prospects\label{sec:detection}}
In this section we study the detection prospects for the proposed scenario
by means of direct and indirect WIMP dark matter experiments.

As can be seen from table~\ref{tb:spectrum} the mass spectrum in this
scenario is quite peculiar. For class A models the LSP is a very pure
bino with a mass in the range $200\;{\rm GeV}\lesssim
m_\chi\lesssim 400$ GeV. The squark and gluino sectors are around $4$
TeV and hence they are essentially decoupled from the neutralino
sector. 
Also the sfermions are heavy, in particular the charginos and the other neutralinos
that are above $1$ TeV. The exception is the lightest stau
$\tilde{\tau}$ that is quite
degenerate in mass with the lightest neutralino.
This implies that the dominant
annihilation channel is $\chi\chi\to\tau^+\tau^-$ through a
$\tilde{\tau}$ exchange and the existence of a very efficient
neutralino-stau coannihilation channel. This latter property make it
possible that the thermal relic density is well below the WMAP
lower bound. From the detection point of view what is
important is the neutralino-neutralino total annihilation cross section. In
the case in hand the dominant annihilation cross section (times velocity $v$) in
$\tau^+\tau^-$ is only of order $10^{-29}\, {\rm cm}^3\, {\rm s}^{-1}$. Due
to the peculiar shape of the energy spectrum of $\gamma$s produced
in the $\tau^+\tau^-$ channel, the class A models can be studied by
carefully analyzing the $\gamma$ ray continuum flux produced by
$\pi^0$ mesons decay coming for example
from the galactic center~\cite{Bergstrom:1997fj}. 
On the other hand the cross sections of the two one-loop processes
$\chi\chi\to\gamma\gamma$ and $\chi\chi\to Z^0\gamma$ are suppressed
by a factor $10^{-2}$ respect to the $\tau^+\tau^-$ cross section and
so are very hard to detect~\cite{Bergstrom:1997fj}.
The direct detection prospects are as well not very encouraging. For
class A models the typical neutralino-proton spin independent cross section is 
\be
\sigma_{\chi p}\simeq 10^{-13}\,{\rm pb}
\ee
well below the current limit of the CDMS
experiment~\cite{Akerib:2005kh} and even below the projected
sensitivity of the upcoming
CDMSII~\cite{Akerib:2006rr}.
The indirect detection through cosmic rays is not favorite as
well. 
To give an estimate we consider the interstellar positron and
antiproton fluxes at fixed energies, $E=10$ GeV for positrons and $E=1$
GeV for antiprotons in order to avoid uncertainties from solar modulation.
Taking into account that all the fluxes scale as $m_\chi^{-2}$
and assuming for the galactic neutralino distribution a NFW profile and the
same propagation model used in~\cite{Baltz:1998xv}
and~\cite{Bergstrom:1999jc}, respectively for the positrons and the
antiprotons, we obtain that the expected interstellar positron flux at
$E=10$ GeV is
around $10^{-11}\,{\rm GeV}^{-1}\,{\rm cm}^{-2}\,{\rm s}^{-1}\,{\rm
  sr}^{-1}$ and the expected interstellar antiproton flux at $E=1$ GeV is around $10^{-7}\,{\rm GeV}^{-1}\,{\rm m}^{-2}\,{\rm s}^{-1}\,{\rm
  sr}^{-1}$ which are well below the experimental limits
determined, for a given
background, in~\cite{Baltz:1998xv,Bergstrom:1999jc}. 
Even with the recently launched PAMELA
satellite~\cite{Picozza:2006nm}, that gives an enhanced resolution and
a better spectrum determination compared with the balloon experiments, the
detection possibilities cannot likely be improved.

In models belonging to the class B the neutralino is a very pure higgsino with $400\;{\rm GeV}\lesssim
m_\chi\lesssim 900$ GeV. From table~\ref{tb:spectrum} we see that while
all the squarks and sleptons are extremely heavy and essentially decoupled
from the lightest neutralino sector
the lightest chargino and the next to lightest neutralino
are quite degenerate in mass with the lightest neutralino. 
The dominant annihilation channels in this case are 
$W^+ W^-$ and $Z^0 Z^0$ with a
cross section (times velocity $v$) of about $10^{-26}\, {\rm
  cm}^3\, {\rm s}^{-1}$. As in the previous case the annihilation cross
section is that required to obtain a thermal relic density in the WMAP
range. The fact that $\Omega_\chi^{{\rm thermal}} h^2<0.09$ is
entirely due to the coannihilations with the second lightest
neutralino and with the lightest chargino. For this class of models
the one-loop processes
$\chi\chi\to\gamma\gamma$ and $\chi\chi\to Z^0\gamma$ gives an
extremely high cross section, with a value of about $2\cdot
10^{-28}\, {\rm cm}^3\, {\rm s}^{-1}$ for a neutralino mass
$m_{\chi}\simeq 400$ GeV. This could be within the reach of the GLAST
sensitivity~\cite{glast-lat-perf}.

From the other side the direct detection is completely out of reach
given the neutralino-proton spin independent cross section 
\be
\sigma_{\chi p}\simeq 10^{-11}\,{\rm pb}
\ee
The cosmic ray signals are expected to be higher respect to the previous
class of models, but still below the sensitivity of PAMELA 
(see for a general analysis~\cite{Profumo:2004ty,Lionetto:2005jd}). 
In fact for class B models the average interstellar positron flux at $E=10$ GeV is
around $10^{-9}\,{\rm GeV}^{-1}\,{\rm cm}^{-2}\,{\rm s}^{-1}\,{\rm
  sr}^{-1}$ while the average interstellar antiproton flux at $E=1$ GeV is
around $5\cdot 10^{-6}\,{\rm GeV}^{-1}\,{\rm m}^{-2}\,{\rm s}^{-1}\,{\rm
  sr}^{-1}$. 

Finally class C models have $1.7\;{\rm GeV}\lesssim
m_\chi\lesssim 2.5$ TeV. The neutralino can be bino-like or mixed. 
The dominant annihilation channel is $b\bar{b}$ for any value of
the gaugino fraction $Z_g$. 
The direct detection rate is favorite in the case of a
mixed-like neutralino, \emph{i.e.} $Z_g=0.5$. The neutralino-proton spin independent cross
section is
\be
\sigma_{\chi p}\simeq 10^{-8}\,{\rm pb}
\ee
so it could be in the reach of CDMSII~\cite{Akerib:2006rr} although it could be very difficult to detect such a high mass neutralino.
Neutralinos with $Z_g=0.5$ or $Z_g\simeq 0.1$ are favorite also in the case of cosmic ray
detection. 
The interstellar positron flux at $E=10$ GeV can be as high as $4\cdot
10^{-9}\,{\rm GeV}^{-1}\,{\rm cm}^{-2}\,{\rm s}^{-1}\,{\rm
  sr}^{-1}$ while the antiproton flux at $E=1$ GeV can be as high as $10^{-4}\,{\rm GeV}^{-1}\,{\rm m}^{-2}\,{\rm s}^{-1}\,{\rm
  sr}^{-1}$ (for a mass $m_\chi \simeq 2.1$ TeV). 
A comparison with the results of~\cite{Baltz:1998xv}
and~\cite{Bergstrom:1999jc} shows that the maximum attainable positron
and antiproton fluxes are
respectively a factor $10^{-2}$ and $10^{-3}$ below the highest
generic MSSM fluxes.  

\section{Conclusions}
We studied the effects of the presence of a single modulus field on the
relic density computation in the the specific framework of the
deflected anomaly mediation. The non thermal contribution to
the relic density allows to resurrect models with $\Omega_\chi^{{\rm thermal}}
h^2$ less than the WMAP lower bound. These models belong to three
different classes with a definite neutralino mass
range and composition and with a peculiar mass spectrum.
The main drawback of this scenario is given by the
fact that there are no models with a very high neutralino pair annihilation
cross section. The low value of the thermal relic density is entirely due to
very efficient coannihilations with particles degenerate in mass 
with the lightest neutralino. As a consequence the detection prospect
of this scenario are not quite promising.

The main result of this analysis is that even invoking a non thermal
production mechanism does not assure in general the existence of
models with high detection rate. The consequences of a non thermal
production has to be carefully analyzed in each specific setup.


\begin{thebibliography}{99}
\bibitem{Spergel:2006hy}
  D.~N.~Spergel {\it et al.}  [WMAP Collaboration],
  arXiv:astro-ph/0603449.
\bibitem{Jungman:1995df}
  G.~Jungman, M.~Kamionkowski and K.~Griest,
  Phys.\ Rept.\  {\bf 267} (1996) 195
  [arXiv:hep-ph/9506380].
\bibitem{Baer:2002fv}
  H.~Baer, C.~Balazs and A.~Belyaev,
  JHEP {\bf 0203} (2002) 042
  [arXiv:hep-ph/0202076].
\bibitem{Moroi:1999zb}
  T.~Moroi and L.~Randall,
  Nucl.\ Phys.\  B {\bf 570} (2000) 455
  [arXiv:hep-ph/9906527].
\bibitem{decaying-scalar}
  D.~J.~H.~Chung, E.~W.~Kolb and A.~Riotto,
  Phys.\ Rev.\  D {\bf 60} (1999) 063504
  [arXiv:hep-ph/9809453].\\
G.~F.~Giudice, E.~W.~Kolb and A.~Riotto,
  Phys.\ Rev.\  D {\bf 64} (2001) 023508
  [arXiv:hep-ph/0005123].\\
R.~Jeannerot, X.~Zhang and R.~H.~Brandenberger,
  JHEP {\bf 9912} (1999) 003
  [arXiv:hep-ph/9901357].\\
W.~B.~Lin, D.~H.~Huang, X.~Zhang and R.~H.~Brandenberger,
  Phys.\ Rev.\ Lett.\  {\bf 86} (2001) 954
  [arXiv:astro-ph/0009003].\\
T.~Moroi, M.~Yamaguchi and T.~Yanagida,
  Phys.\ Lett.\  B {\bf 342} (1995) 105
  [arXiv:hep-ph/9409367].\\
M.~Kawasaki, T.~Moroi and T.~Yanagida,
  Phys.\ Lett.\  B {\bf 370} (1996) 52
  [arXiv:hep-ph/9509399].\\
S.~Khalil, C.~Munoz and E.~Torrente-Lujan,
  New J.\ Phys.\  {\bf 4} (2002) 27
  [arXiv:hep-ph/0202139].\\
N.~Fornengo, A.~Riotto and S.~Scopel,
  Phys.\ Rev.\  D {\bf 67} (2003) 023514
  [arXiv:hep-ph/0208072].\\
M.~Endo, K.~Hamaguchi and F.~Takahashi,
  Phys.\ Rev.\ Lett.\  {\bf 96} (2006) 211301
  [arXiv:hep-ph/0602061].\\
S.~Nakamura and M.~Yamaguchi,
  Phys.\ Lett.\  B {\bf 638} (2006) 389
  [arXiv:hep-ph/0602081].
\bibitem{Silverstein:2004id}
  E.~Silverstein,
  arXiv:hep-th/0405068.
\bibitem{Banks:1995dt}
  T.~Banks, M.~Berkooz and P.~J.~Steinhardt,
  Phys.\ Rev.\  D {\bf 52} (1995) 705
  [arXiv:hep-th/9501053].
\bibitem{Rattazzi:1999qg}
  A.~Pomarol and R.~Rattazzi,
  JHEP {\bf 9905} (1999) 013
\bibitem{Okada:2002mv}
N.~Okada,
Phys.\ Rev.\ D {\bf 65} (2002) 115009
\bibitem{Cesarini:2006jp}
  A.~Cesarini, F.~Fucito and A.~Lionetto,
  Phys.\ Rev.\  D {\bf 75} (2007) 025026
  [arXiv:hep-ph/0611098].
\bibitem{Kolb:1988aj}
  E.~W.~.~Kolb and M.~S.~.~Turner,
  ``THE EARLY UNIVERSE. REPRINTS,''
{\it  REDWOOD CITY, USA: ADDISON-WESLEY (1988) 719 P. (FRONTIERS IN
  PHYSICS, 70)}
\bibitem{Giudice:1998bp}
G.~F.~Giudice and R.~Rattazzi,
Phys.\ Rept.\  {\bf 322} (1999) 419
\bibitem{Gondolo:2004sc}
  P.~Gondolo, J.~Edsjo, P.~Ullio, L.~Bergstrom, M.~Schelke and E.~A.~Baltz,
  JCAP {\bf 0407} (2004) 008
  [arXiv:astro-ph/0406204].
\bibitem{Akerib:2005kh}
  D.~S.~Akerib {\it et al.}  [CDMS Collaboration],
  Phys.\ Rev.\ Lett.\  {\bf 96} (2006) 011302
  [arXiv:astro-ph/0509259].
\bibitem{Akerib:2006rr}
  D.~S.~Akerib {\it et al.},
  Nucl.\ Instrum.\ Meth.\  A {\bf 559} (2006) 411.
\bibitem{Bergstrom:1997fj}
  L.~Bergstrom, P.~Ullio and J.~H.~Buckley,
  Astropart.\ Phys.\  {\bf 9} (1998) 137
  [arXiv:astro-ph/9712318].
\bibitem{Baltz:1998xv}
  E.~A.~Baltz and J.~Edsjo,
  Phys.\ Rev.\  D {\bf 59} (1999) 023511
  [arXiv:astro-ph/9808243].
\bibitem{Bergstrom:1999jc}
  L.~Bergstrom, J.~Edsjo and P.~Ullio,
  Astrophys.\ J.\  {\bf 526} (1999) 215
  [arXiv:astro-ph/9902012].
\bibitem{Picozza:2006nm}
  P.~Picozza {\it et al.},
  Astropart.\ Phys.\  {\bf 27} (2007) 296
  [arXiv:astro-ph/0608697].
\bibitem{glast-lat-perf}
\url{http://www-glast.slac.stanford.edu/software/IS/glast_lat_performance.htm}
\bibitem{Profumo:2004ty}
  S.~Profumo and P.~Ullio,
  JCAP {\bf 0407} (2004) 006
  [arXiv:hep-ph/0406018].
\bibitem{Lionetto:2005jd}
  A.~M.~Lionetto, A.~Morselli and V.~Zdravkovic,
  JCAP {\bf 0509} (2005) 010
  [arXiv:astro-ph/0502406].
\end{thebibliography}
\end{document}